\documentclass{article}
\RequirePackage{amsmath} 

\usepackage[compatibility=false]{caption}

\usepackage{booktabs} 
\usepackage{subcaption} 
\usepackage{graphicx}
\usepackage{pgfplots}
\usepackage[all]{nowidow}
\usepackage[utf8]{inputenc}
\usepackage{tikz}
\usetikzlibrary{er,positioning,bayesnet}
\usepackage{multicol}
\usepackage{algpseudocode,algorithm,algorithmicx}

\usepackage{hyperref}
\usepackage[inline]{enumitem} 

\definecolor{blue}{HTML}{1F77B4}
\definecolor{orange}{HTML}{FF7F0E}
\definecolor{green}{HTML}{2CA02C}

\pgfplotsset{compat=1.14}

\usepackage[numbers,round]{natbib} 
\begin{document}
\title{Variability in Aggregate Personal Income Across Industrial Sectors During COVID-19 Shock: A Time-Series Exploration}
%
%
\author{Didarul Islam , Mohammad Abdullah Al Faisal}
%
%

%
\maketitle              
\begin{abstract}
{This study explored the variability in Aggregate Personal Income (PI) across 13 major industrial sectors in the US during the COVID-19 pandemic. Utilizing time-series data from 2010 Q1 to 2019 Q4, we employed Autoregressive Integrated Moving Average (ARIMA) models to establish baseline trends in Personal Income (PI) before the pandemic. We then extended these models to forecast PI values for the subsequent 14 quarters, from 2020 Q1 to 2023 Q2, as if the pandemic had never happened. This forecasted data was compared with the actual PI data collected during the pandemic to quantify its impacts. This approach allowed for the assessment of both immediate and extended effects of COVID-19 on sector-specific PI. Our study highlighted the resilience of PI in sectors like Utilities, Retail, Finance, Real Estate, and Healthcare, with Farming showing an early recovery in PI, despite significant initial setbacks. In contrast, PI in Accommodation and Food Services experienced delayed recovery, contributing significantly to the overall impact variance alongside Farming (53.26\% and 33.26\% respectively). Finance and Utilities demonstrated positive deviations, suggesting a lesser impact or potential benefit in early pandemic stages. Meanwhile, sectoral PI in Manufacturing, Wholesale and Education showed moderate recovery, whereas Construction and Government lagged in resilience. The aggregate economic impact, initially negative at -0.027 in 2020 Q1, drastically worsened to -1.42 in Q2, but improved by Q4, reflecting a broader trend of adaptation and resilience across all the sectors during the pandemic.}
\end{abstract}
\section{Introduction}
The COVID-19 pandemic leaves its footprint in many areas, and most of these ramifications are negative in nature. Reduction in wages and Job-loss is one of the most adverse effects of it, creating many more other problems. According to Reference \cite{shrider2021income}, the unemployment rate hit 14.7\%, making it the highest employment rate in April 2020, forcing 20 percent of all workers to request unemployment benefits before July 2020 in the United States \cite{donnelly2021state}. The cutback in employment occurred in two ways: employers cut short some of their employees’ working hours, and some had to take care of their sick family members, which was mostly provided by mothers \cite{shrider2021income, brenan2020us, modestino2020coronavirus}. Reference \cite{cajner2020us} reported that two-thirds of existing employed workers faced such a decrease in income. Even though it was not distressing for high-income workers, low-income and middle-income workers were highly affected by it. Moreover, not only was this decline in income different for different races \cite{anyamele2021disparities} but also it was different for the different sectors of the US economy. This paper explores such variabilities in income across Industrial Sectors by examining the before and after the pandemic trend of aggregate Personal Income (PI).

The COVID-19 pandemic significantly impacted global economies, necessitating analysis of its effect on key economic indicators like Personal Income (PI). This study examines PI variability across 13 major US industry sectors, contributing to about 90\% of the economy, including Farming (FARM), Utilities (UTL), Construction (CONS), manufacturing (MAN), Wholesale (WHO), Retail (RET), Transportation (TRANS), Finance (FIN), Real Estate (RE), Education (EDU), Health (HEA), Accommodation and Food Services (ACCF), and Government (Gov).

In the first section, this paper points out the disparities in income across different groups of people by citing different studies and reports, and it also outlines the need for this paper to adopt appropriate policies. Section 2 summarizes the contemporary literature on this topic. Data description is in section 3, and section 4 discusses the methodology and findings. Finally, section 5 concludes the paper.

\section{Literature Review}
Given the recent emergence of the COVID-19 pandemic, the body of literature related to this topic is relatively limited. However, many scholars have studied the phenomenon from various aspects to restore the world's economic status and the health status of people. This section summarizes such studies. 

There were two more main coronaviruses (SARS-CoV in 2002 and MARS-CoV in 2012) before COVID-19, but they could not spread as much as the COVID-19 pandemic did. As a result, the economic and health consequences of COVID-19 were more damaging and ubiquitous than those of the other two outbreaks. Reference \cite{deaton2022gdp} disentangled these economic and health effects, and they argued that health should be prioritized over material well-being. They provided an example to support their claim pointing out the fact that the expenditure on health does not say anything about the outcomes on health \citep{lau2003monitoring, deaton2022gdp}. Reference \citep{difusco2021health} reported their results in terms of both health and economic outcomes. Reference \cite{donnelly2021state} examined the nexus between job loss and mental health. They found that the effects of being unemployed are different for every individual. Some were more anxious and stressed than others \cite{donnelly2021state}.
Reference \cite{dang2021gender} explored the effects of COVID-19 on the gender gap. They claimed that women were more likely to be unemployed than men \cite{dang2021gender}. Moreover, reference \cite{anyamele2021disparities} extended their search and investigated the effects of COVID-19 on the loss of jobs and the reduction of income in the context of race, gender, and different age groups in the United States by employing multivariate logistic regression and Oaxaca-Blinder decomposition analysis. They concluded that its influence was more intense on Blacks, Asians, Hispanics, and others than on Whites \cite{anyamele2021disparities}. Almost similar findings were discovered by \cite{hall2022income}. They claimed that COVID-19 affected people according to their financial status and societal status. Low-income people and minorities were more vulnerable \cite{hall2022income}. Reference \cite{egger2021falling} argued that this susceptibility was not limited to the individual level. A sharp decline in income, employment, and food security across low-income and middle-income countries during the COVID crisis was found by \cite{egger2021falling}. On the other hand, reference \cite{han2020income} found that such consequences of COVID-19 were mitigated by the policies adopted by the government \cite{han2020income}. Instead of researching the effects of COVID-19 on income among different ethnic groups, reference \cite{hanspal2020income} explored its impacts on income and wealth. They concluded that wealth shocks were more robust among middle-aged and elderly people while income shocks were stronger among young people \cite{hanspal2020income}.
Unable to move from one place to another caused a devastating crisis in the United States and the world \cite{anyamele2021disparities}. The findings of \cite{delrio2020supply} supported the claim of  \cite{anyamele2021disparities}. They found that the industries which had the opportunity to work remotely were less affected while industries like transportation, tourism, and entertainment faced serious supply and demand shocks \cite{delrio2020supply}. Conversely, reference \cite{goolsbee2021fear} argued that the fear of being infected was a more crucial factor leading to such an economic crisis than government restriction on individual movement and business operations \cite{goolsbee2021fear}. However, not everybody obeys the limitations imposed by the government. Reference \cite{lau2003monitoring} claims that educated and high-income people are more obedient to such laws than others. Reference \cite{pew2020worries} also argued that people with college degrees were more likely to acknowledge the coronavirus as a threat to society than people who did not have any college degrees. 
COVID-19 is an unprecedented event, which has caused the death of millions of people, either through health complications or economic hardship. Many scholars have already studied this phenomenon from different angles, and many are still exploring the causes and remedies of its effects. However, to the best of researchers’ knowledge, no studies have been done focusing on PI of the Industrial Sector specifically. This paper aims to fill that research gap by concentrating on sector-specific impacts, which adds a new dimension to COVID-19 economic studies, offering a detailed understanding of the pandemic’s ramifications on PI of the Industrial Sectors and informing targeted policy interventions. Our analysis is crucial for understanding sector resilience and recovery during the crisis, aiding in economic policy formulation and resource allocation. By filling this research gap, our study not only adds a novel dimension to the understanding of COVID-19's economic impacts but also highlights the importance of sector-specific analysis in comprehending the bigger picture of a global crisis's effects on aggregate personal income. This distinctive focus makes our research both important and relevant for future economic planning and recovery efforts.

\section{Data Description}
The dataset utilized for this research is sourced from the BEA (Bureau of Economic Analysis) regional data website. It consists of aggregate Personal Income (PI) data, log-transformed for analytical robustness, spanning from the first quarter of 2010 (2010 Q1) to the second quarter of 2023 (2023 Q2). The data comprehensively covers 13 major industry sectors in the United States, which collectively represent about 90\% of all industry sectors. These sectors are crucial for examining the variability of personal income in the face of pandemic.
Log transformation of the data aids in normalizing the distribution and reducing the effect of outliers, thus providing a more accurate representation of trends and patterns. The summary statistics for the log-transformed PI data across these sectors are as follows:
\begin{table}[ht]
\centering
\begin{tabular}{|l|c|c|c|}
\hline
\textbf{Sector} & \textbf{Minimum} & \textbf{Mean} & \textbf{Maximum} \\ \hline
FARM            & 17.82            & 18.29         & 18.72            \\ \hline
UTL             & 18.06            & 18.38         & 18.72            \\ \hline
CONS            & 19.99            & 20.33         & 20.69            \\ \hline
MAN             & 20.56            & 20.80         & 21.06            \\ \hline
WHO             & 19.88            & 20.14         & 20.45            \\ \hline
RET             & 20.08            & 20.33         & 20.62            \\ \hline
TRANS           & 19.51            & 19.91         & 20.33            \\ \hline
FIN             & 20.29            & 20.52         & 20.84            \\ \hline
RE              & 18.76            & 19.41         & 19.81            \\ \hline
EDU             & 18.84            & 19.12         & 19.41            \\ \hline
HEA             & 20.72            & 20.98         & 21.32            \\ \hline
ACCF            & 19.37            & 19.76         & 20.19            \\ \hline
GOV             & 21.23            & 21.37         & 21.57            \\ \hline
\end{tabular}
\caption{Summary Statistics of Personal Income Across Sectors}
\label{tab:my_label}
\end{table}

\section{Methodology and Findings}
This study employed time-series analysis to examine the variability in Personal Income (PI) across various industry sectors during Covid 19. To capture the pre-pandemic trends and isolate the impact of COVID-19, ARIMA models were fitted to data up to 2019 Q4 (end of the pre-COVID period) for each industry. This approach allowed for a baseline against which to compare the pandemic period's deviations.\\
\begin{table}[ht]
\centering
\begin{tabular}{@{}lllr@{}}
\toprule
Industry & ARIMA(p,d,q) & Model Description & BIC \\ \midrule
FARM     & ARIMA(0,1,0) & Random walk & -38.88 \\
UTL      & ARIMA(1,1,0) with drift & AR with drift & -195.07 \\
CONS     & ARIMA(0,1,0) with drift & Random walk with drift & -233.73 \\
MAN      & ARIMA(0,1,0) with drift & Random walk with drift & -244.79 \\
WHO      & ARIMA(0,1,0) with drift & Random walk with drift & -238.21 \\
RET      & ARIMA(0,1,1) with drift & MA with drift & -261.7 \\
TRANS    & ARIMA(0,1,0) with drift & Random walk with drift & -235.13 \\
FIN      & ARIMA(1,1,0)(0,0,1)[4] with drift & AR and seasonal MA with drift & -173.17 \\
RE       & ARIMA(0,2,1) & MA with double differencing & -177.43 \\
EDU      & ARIMA(0,1,0)(0,0,1)[4] with drift & Seasonal MA with drift & -274.44 \\
HEA      & ARIMA(0,1,0) with drift & Random walk with drift & -296.26 \\
ACCF     & ARIMA(0,1,0) with drift & Random walk with drift & -267.37 \\
GOV      & ARIMA(1,2,1) & AR and MA with double differencing & -300.09 \\
\bottomrule
\end{tabular}
\caption{Summary of ARIMA Model Findings for Various Industries}
\label{tab:arima_models}
\end{table}

ARIMA models, known for their flexibility in handling various time series patterns, were individually tailored for each sector. The model selection process, guided by the auto.arima function, determined the optimal combination of autoregressive (AR), differencing (I), and moving average (MA) components for each time series.  The general form of an ARIMA model is ARIMA(p,d,q), where 'p' is the number of AR terms, 'd' is the degree of differencing, and 'q' is the number of MA terms.
A basic ARIMA model equation can be represented as:
$$
\left(1-\sum_{i=1}^p \phi_i L^i\right)(1-L)^d X_t=\left(1+\sum_{j=1}^q \theta_j L^j\right) \epsilon_t
$$

Where:\\
$X_t$ is the time series data.\\
$\phi_i$ are the parameters of the AR terms.\\
$L$ is the lag operator.\\
$d$ is the order of differencing.\\
$\theta_j$ are the parameters of the MA terms.\\
$\epsilon_t$ is white noise error.\\
The summary of the ARIMA
findings are listed in table 2.
\\
In our methodology, we employed a two-step process to analyze the impact of COVID-19 on various industries.
First, we forecasted Personal Income (PI) for each industry using ARIMA models, projecting 14 quarters ahead from the first quarter of 2020, based on data up to the end of 2019. This forecasting created a baseline to compare against the actual PI observed during the pandemic. We selected a specific window of the time series data, from 2020 Q1 to 2023 Q2, to focus on the period most likely to be impacted by COVID-19. This window includes both the onset of the pandemic and subsequent quarters, allowing for an analysis of both immediate and longer-term effects.
Next, we calculated the impact of the pandemic on each industry by comparing these forecasted values with the actual data from 2020 Q1 to 2023 Q2. This comparison was made for each quarter, allowing us to assess both the immediate and extended impacts of the pandemic. The impact was quantified as the difference between the forecasted (expected without COVID-19) and actual PI values across all 13 sectors.

\subsection{Aggregate Shock}
Table 3 depicts a timeline of the aggregated economic impact of COVID-19 on PI from the first quarter of 2020 through the second quarter of 2023. Initially, in 2020 Q1, the impact was slightly negative at -0.027, which drastically worsened in Q2 at-1.42, reflecting the immediate and severe effect of the pandemic's onset. Subsequent quarters in 2020 showed a gradual recovery, turning positive in Q4, suggesting adaptation and resilience in various industries.

The year 2021 and beyond showed predominantly positive aggregate impacts, indicating a consistent recovery trajectory across industries. This trend continued into 2023, with all quarters showing positive aggregate impacts.This suggests that the overall Personal Income didn't just bounce back from the initial shock, but it also grew more than what we might have expected if the pandemic had never happened.

\begin{table}[h]
\centering
\label{tab:aggregate_impact}
\begin{tabular}{@{}lr@{}}
\toprule
Quarter & Aggregate Impact \\ \midrule
2020 Q1 & -0.02743556 \\
2020 Q2 & -1.42218146 \\
2020 Q3 & -0.31169539 \\
2020 Q4 &  0.23587149 \\
2021 Q1 & -0.04904149 \\
2021 Q2 &  0.69007384 \\
2021 Q3 &  0.75464512 \\
2021 Q4 &  0.59913413 \\
2022 Q1 &  0.61505964 \\
2022 Q2 &  0.64900868 \\
2022 Q3 &  0.75491565 \\
2022 Q4 &  0.71369990 \\
2023 Q1 &  0.72154733 \\
2023 Q2 &  0.61180581 \\ \bottomrule
\end{tabular}
\caption{Aggregated Impact of COVID-19 on Personal Income Across Industries}
\end{table}

\subsection{Sectoral Shock Analysis}
Table 4 in the study shows the shift from negative to positive impacts on Personal Income (PI) across 13 sectors during the COVID-19 pandemic. It categorizes sectors into four resilience levels based on their recovery timing and magnitude.\\

\begin{table}[h]
\centering
\label{tab:industry_resilience_sorted}
\begin{tabular}{@{}lllr@{}}
\toprule
Sector & Turning Point Quarter & Impact Value & Resilience Category \\ \midrule
\textbf{Highly Resilient:} \\
FARM   & 2020 Q3               & 0.0159       & Highly Resilient \\
FIN    & 2020 Q1               & 0.0468       & Highly Resilient \\
HEA    & 2020 Q3               & 0.015        & Highly Resilient \\
RE     & 2020 Q4               & 0.011        & Highly Resilient \\
RET    & 2020 Q3               & 0.0032       & Highly Resilient \\
UTL    & 2020 Q1               & 0.0295       & Highly Resilient \\
\textbf{Moderately Resilient:} \\
TRANS  & 2021 Q2               & 0.0083       & Moderately Resilient \\
\textbf{Less Resilient:} \\
ACCF   & 2021 Q3               & 0.0511       & Less Resilient \\
EDU    & 2021 Q4               & 8.5e-5       & Less Resilient \\
MAN    & 2021 Q4               & 0.0013       & Less Resilient \\
WHO    & 2021 Q4               & 0.0015       & Less Resilient \\
\textbf{Minimally Resilient:} \\
CONS   & 2023 Q2               & -0.025       & Minimally Resilient \\
GOV    & 2023 Q2               & -0.0149      & Minimally Resilient \\
\bottomrule
\hline
\multicolumn{4}{l}{\textbf{Resilience Category Definitions:}} \\
\multicolumn{4}{l}{\textbf{Highly Resilient:} Recovery in 2020 Q1 - 2020 Q4.} \\
\multicolumn{4}{l}{\textbf{Moderately Resilient:} Recovery in 2021 Q1 - 2021 Q2.} \\
\multicolumn{4}{l}{\textbf{Less Resilient:} Recovery in 2021 Q3 - 2022 Q1.} \\
\multicolumn{4}{l}{\textbf{Minimally Resilient:} Recovery beyond 2022 Q1 or no recovery.} \\
\hline
\end{tabular}
\caption{Summary of Turning Points from Negative to Positive Impact (Sorted)}
\end{table}
Agriculture, Utilities, Retail, Finance, Real Estate, and Healthcare, demonstrating rapid PI recovery between 2020 Q1-Q4, are classified as 'Highly Resilient.' The Transportation sector, rebounding between 2021 Q1-Q2, is 'Moderately Resilient.' Conversely, Manufacturing, Wholesale, Education, and Accommodation and Food Services, recovering later (2021 Q3-2022 Q1), are 'Less Resilient.' Notably, Construction and Government sectors, either with slow recovery post-2022 Q1 or no recovery within the study's scope, are 'Minimally Resilient.' This signifies the varied resilience and adaptability of sector-specific PI in response to the pandemic's economic impact.
The graphs outline the variability of Personal Income (PI) across different industry sectors over time, from the first quarter of 2020 to the second quarter of 2023 in response to Covid-19 shock. The impacts are measured as deviations from the forecasted PI without the pandemic.The key findings include:\\
\begin{itemize}
    \item The Farm sector experienced significant volatility, with substantial negative impacts at the beginning of the pandemic, followed by a recovery, and another dip towards the end of the period studied.
    \item The Utility sector shows the least variability and remains relatively stable with minor positive and negative deviations.
    \item The Construction sector had a moderate negative impact initially but shows a recovery trend with minimal fluctuations thereafter.
    \item The Manufacturing sector was affected initially too but bounced back in 2020 Quarter 4, suggesting a less negative or even potentially positive impact on PI in this sector.
    \end{itemize}

\begin{figure}[h]
  \centering
\includegraphics[width=0.8\textwidth]{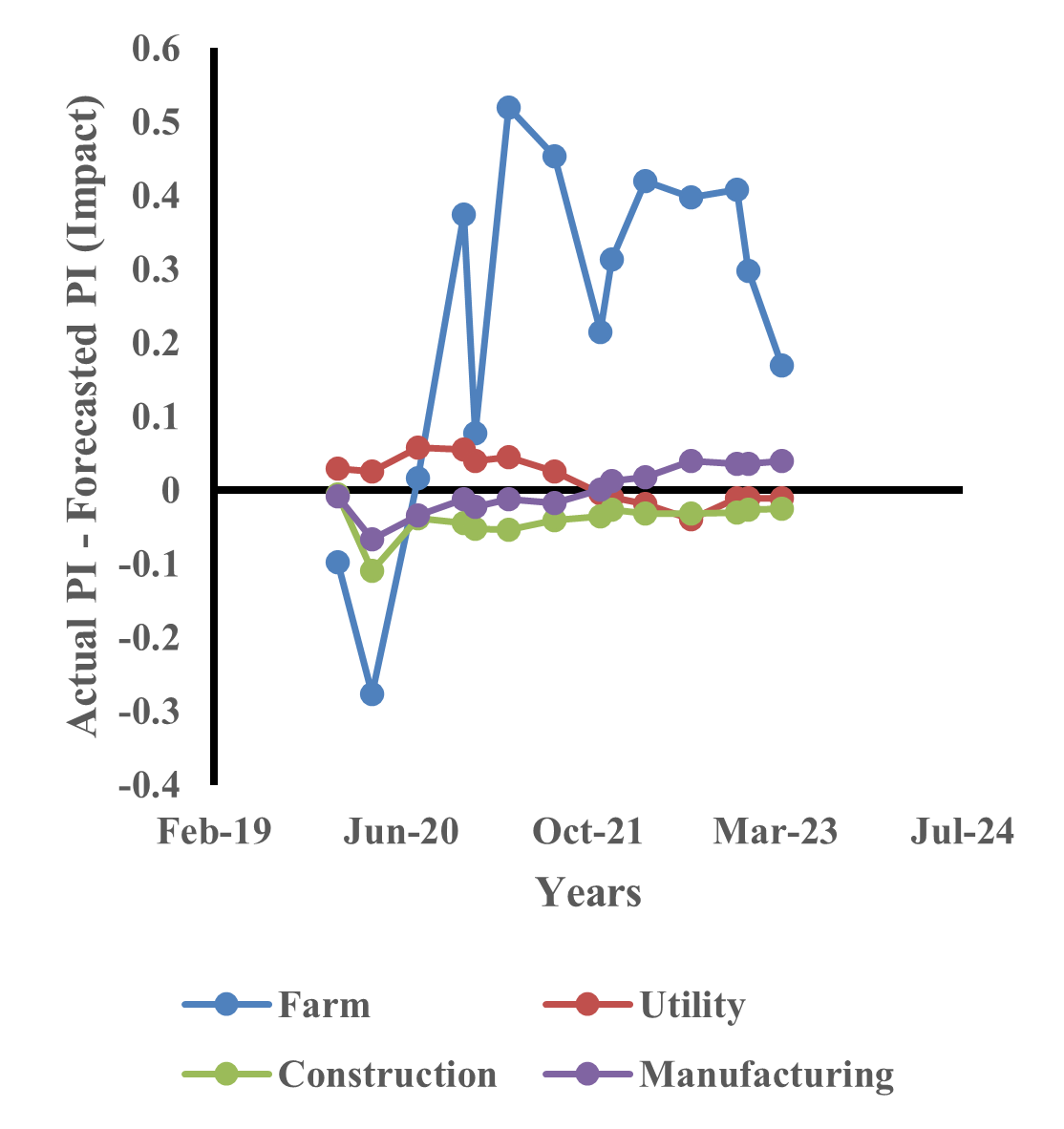}
  \caption{Impact on industries over quarters}
  \label{fig:graph1}
\end{figure}

\begin{figure}[h]
  \centering
  \includegraphics[width=0.8\textwidth]{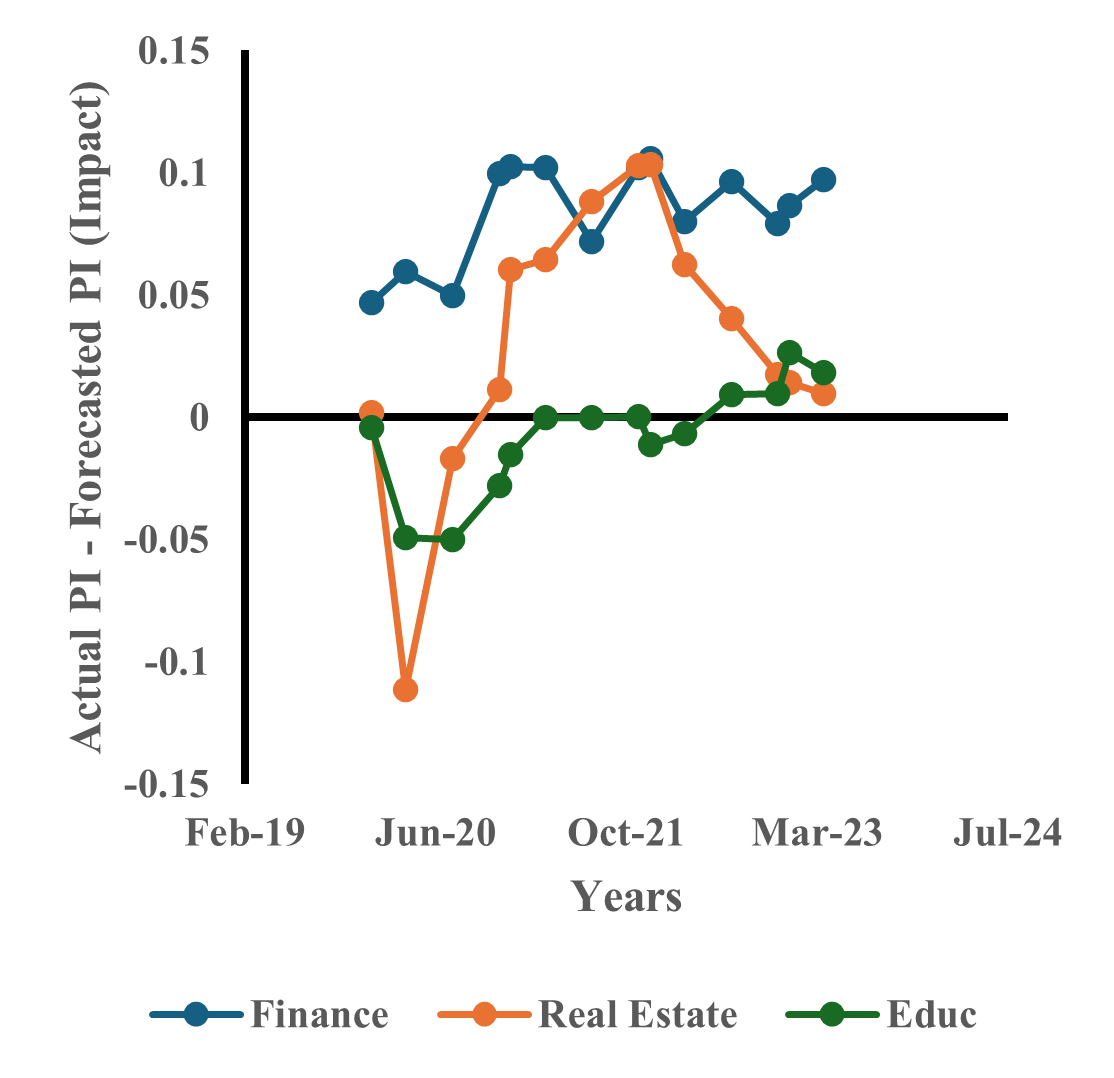}
  \caption{Impact on industries over quarters}
  \label{fig:graph2}
\end{figure}

\begin{figure}[h]
  \centering
\includegraphics[width=0.8\textwidth]{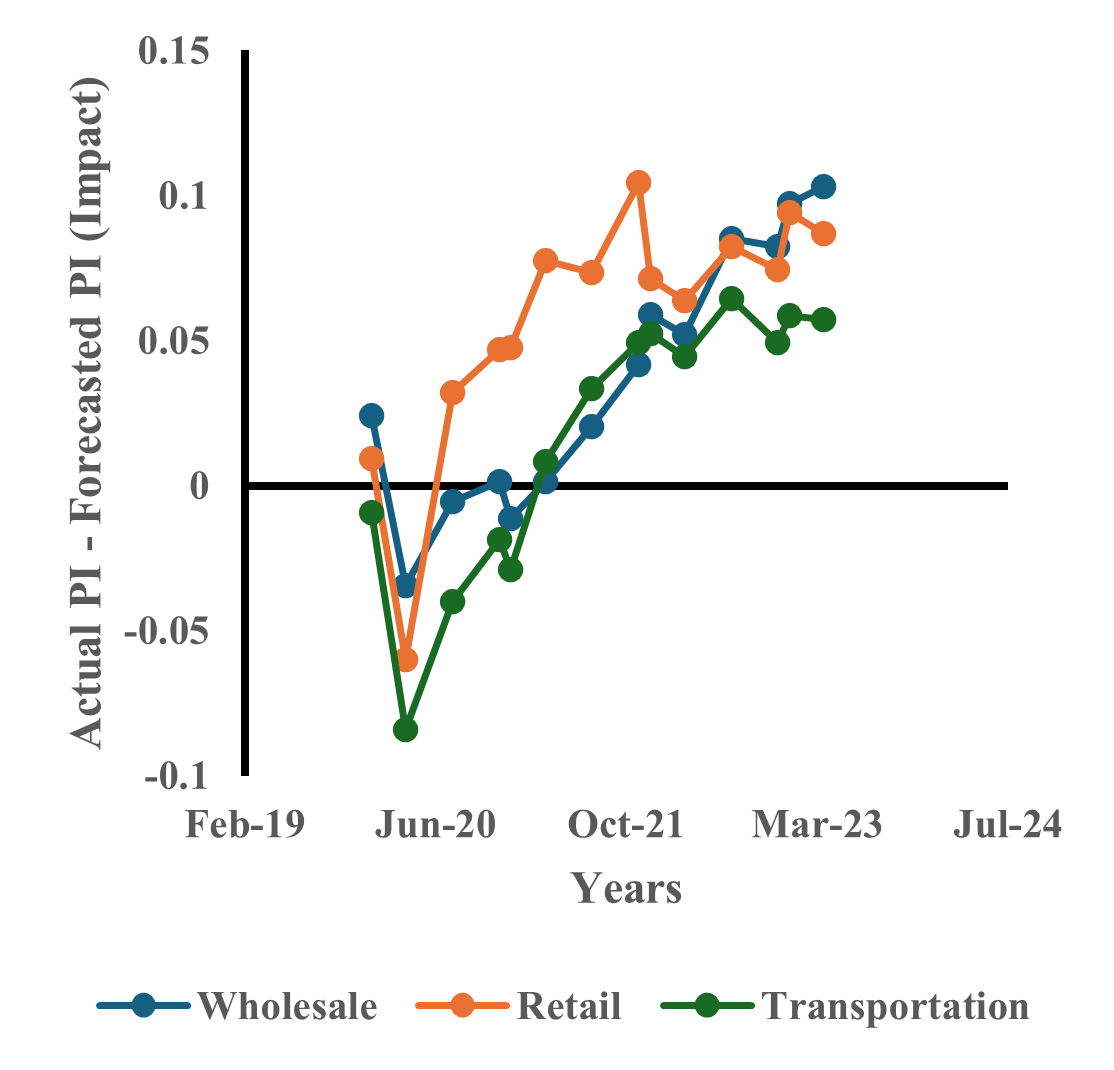}
  \caption{Impact on industries over quarters}
  \label{fig:graph3}
\end{figure}
\begin{figure}[H]
  \centering
\includegraphics[width=0.8\textwidth]{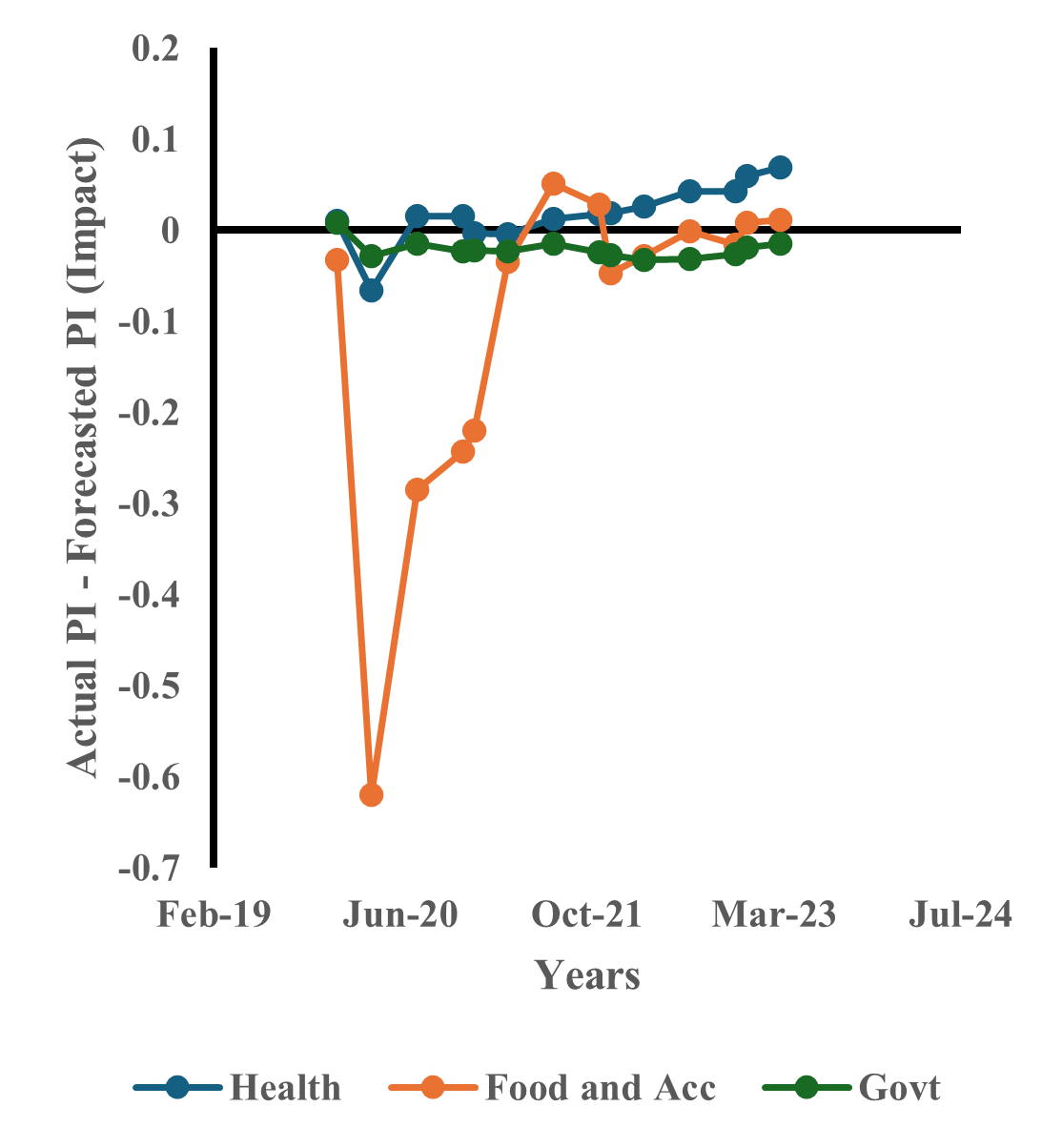}
  \caption{Impact on industries over quarters}
  \label{fig:graph4}
\end{figure}

    \begin{itemize}
    \item The Wholesale sector saw an initial sharp dip in the face of the pandemic, but it was able to bounce back quickly and remained stable afterwards.
    \item The Retail sector faced a sharp decline at the start of the pandemic, followed by a swift recovery and eventual stabilization, with minor fluctuations above and below the baseline.
    \item The Transportation sector experienced the most significant initial drop, indicating a strong immediate impact, but then rebounded and fluctuated around the forecasted baseline.
    \item The Finance sector shows resilience, with mostly positive deviations, suggesting that PI in this sector was less affected or potentially benefited during the pandemic.
    \item The Real Estate sector exhibits a substantial negative impact initially, then a recovery with fluctuations close to the forecasted values.
    \item The Education sector (labeled as "Educ") experienced a significant initial shock with a deep decline in PI, followed by a gradual recovery, but still below the baseline for most of the period.
    \item The Health sector had a slight decline in PI at the beginning of the pandemic, but it quickly recovered and remained stable, even experiencing some positive deviations from the forecasted PI.
    \item The Food and Accommodation sector (labeled as "Food and Acc") shows a severe initial impact with a steep drop in PI. However, it demonstrates a recovery trend, approaching the baseline over time.
    \item The Government sector (labeled as "Govt") experienced minimal impact, with slight fluctuations around the baseline, indicating stability in PI during the pandemic.
\end{itemize}
\begin{figure}[h]
  \centering
\includegraphics[width=0.8\textwidth]{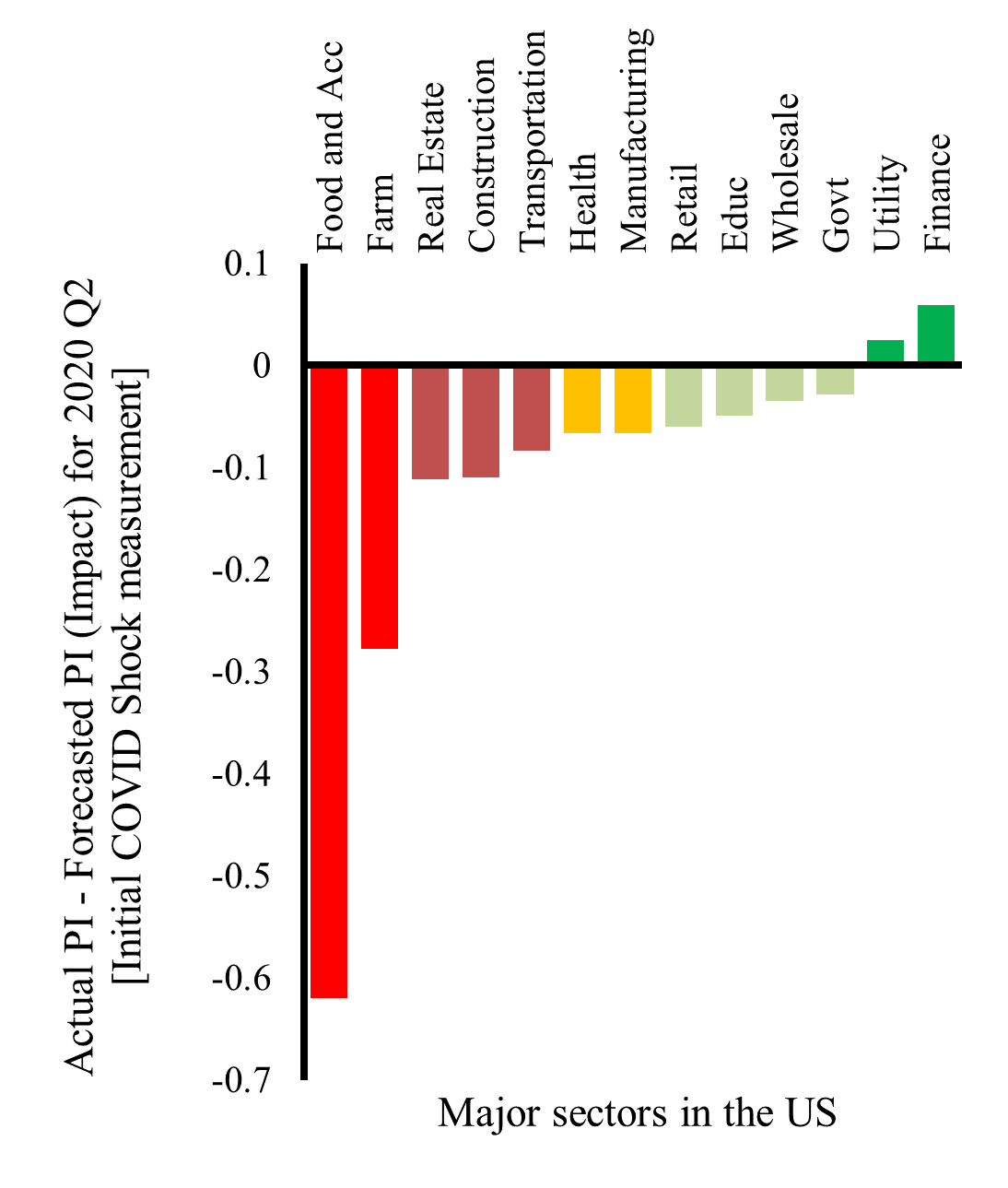}
  \caption{Initial Industry Specific PI Response to Shock}
  \label{fig:graph5}
\end{figure}

The magnitude of initial shock in PI across various sectors can be seen from figure 4. The graph provides a snapshot of the impact of the COVID-19 pandemic on Personal Income (PI) across various industry sectors for the second quarter of 2020, when the impact of the shock was peaked. The x-axis represents the different sectors, the y-axis indicates the magnitude of deviation from the forecasted PI, where the horizontal red line suggests the baseline (zero deviation) forecasted without the pandemic's impact.\\

Key findings include:\\
\begin{itemize}
    \item The ``Food and Accommodation" sector and ``Farm" appear to have experienced the most significant initial negative impact, with PI falling far below the forecasted levels.
    \item Finance and Utility sectors show a positive deviation, indicating that PI was higher than expected despite the pandemic. These two industries showed strong resilience to the initial shock of the pandemic.
    \item Other sectors like ``Construction," ``Health," ``Transportation," ``Real Estate," and ``Manufacturing" show negative impacts but to a lesser extent than the ``Food and Accommodation" or Farm sector.
    \item The ``Wholesale," ``Retail," ``Education," and ``Govt" sectors are close to the expected line, suggesting a minimal impact from the pandemic during the initial period.
\end{itemize}
\subsection{Sectoral contributions to Impact Variance}

The analysis of the variance in the impact of COVID-19 across various industries reveals significant disparities. The data table showcases each industry's contribution to the overall variance in impact. Notably, the `FARM' and `ACCF' industries exhibit the highest variances, contributing 53.26\% and 33.26\% respectively to the total variance. This suggests that these sectors experienced the most significant fluctuations in impact due to the pandemic. In contrast, industries like `UTL', `CONS', and `FIN' show considerably lower variance contributions, indicating more uniform impacts within these sectors. 

\begin{table}[ht]
\centering
\label{tab:impact_variance}
\begin{tabular}{lrr}
\hline
\textbf{Industry} & \textbf{Impact Variance} & \textbf{Contribution (\%)} \\ \hline
FARM  & 0.05378 & 53.26 \\
UTL   & 0.00094 & 0.93  \\
CONS  & 0.00055 & 0.55  \\
MAN   & 0.00099 & 0.98  \\
WHO   & 0.00195 & 1.93  \\
RET   & 0.00179 & 1.77  \\
TRANS & 0.00212 & 2.10  \\
FIN   & 0.00041 & 0.41  \\
RE    & 0.00319 & 3.16  \\
EDU   & 0.00051 & 0.50  \\
HEA   & 0.00106 & 1.05  \\
ACCF  & 0.03359 & 33.26 \\
GOV   & 0.00011 & 0.11  \\ \hline
\end{tabular}
\caption{Contribution of Each Industry to the Total Variance of Impact}
\end{table}

\section{Conclusion and Policy}
This study adopted a unique approach involved forecasting Personal Income (PI) trends as if the COVID-19 pandemic had never occurred, providing a novel baseline to measure the actual pandemic's impact against. This method allowed for an intriguing comparison of `what-if' scenarios, offering a distinctive perspective on the pandemic's true economic imprint across various sectors. Our findings indicate that while sectors like Accommodation and Food Services, and Farming faced significant initial setbacks, others such as Utilities and Finance demonstrated remarkable stability and even positive growth. This suggests a nuanced economic impact of the pandemic, varying significantly across sectors. Policymakers should consider these insights for targeted economic support, focusing on sectors that lag in recovery while leveraging the strengths of more resilient industries like finance and utilities. Future policies could benefit from a sector-specific approach, ensuring efficient allocation of resources and support where it is most needed to foster a balanced and robust economic recovery.


\appendix

\section{Appendix}
\subsection{Sectoral Impact Analysis Tables}
\begin{table}[H]
\centering
\begin{tabular}{|c|c|c|c|}
\hline
Quarter & Forecast & Actual & Impact \\
\hline
1 & 18.16851 & 18.07062 & -0.09789289 \\
2 & 18.16851 & 17.89125 & -0.27726308 \\
3 & 18.16851 & 18.18444 & 0.01592916 \\
4 & 18.16851 & 18.54236 & 0.37385135 \\
5 & 18.16851 & 18.24560 & 0.07709573 \\
6 & 18.16851 & 18.68773 & 0.51922607 \\
7 & 18.16851 & 18.62178 & 0.45327377 \\
8 & 18.16851 & 18.38392 & 0.21541231 \\
9 & 18.16851 & 18.48193 & 0.31342446 \\
10 & 18.16851 & 18.58851 & 0.42000088 \\
11 & 18.16851 & 18.56654 & 0.39803591 \\
12 & 18.16851 & 18.57592 & 0.40741426 \\
13 & 18.16851 & 18.46649 & 0.29797683 \\
14 & 18.16851 & 18.33809 & 0.16957942 \\
\hline
\end{tabular}
\caption{Impact on the FARM industry}
\label{tab:impact_farm}
\end{table}

\begin{table}[H]
\centering
\begin{tabular}{|c|c|c|c|}
\hline
Quarter & Forecast & Actual & Impact \\
\hline
1 & 18.56681 & 18.59627 & 0.029465380 \\
2 & 18.58017 & 18.60524 & 0.025071353 \\
3 & 18.59310 & 18.65116 & 0.058060590 \\
4 & 18.60590 & 18.66070 & 0.054795887 \\
5 & 18.61867 & 18.65776 & 0.039083191 \\
6 & 18.63143 & 18.67577 & 0.044342557 \\
7 & 18.64419 & 18.66939 & 0.025203604 \\
8 & 18.65694 & 18.65236 & -0.004582726 \\
9 & 18.66970 & 18.66072 & -0.008975215 \\
10 & 18.68245 & 18.66427 & -0.018177012 \\
11 & 18.69521 & 18.65578 & -0.039423099 \\
12 & 18.70796 & 18.69664 & -0.011316736 \\
13 & 18.72071 & 18.70939 & -0.011323244 \\
14 & 18.73347 & 18.72283 & -0.010634483 \\
\hline
\end{tabular}
\caption{Impact on the Utility industry}
\label{tab:impact_utl}
\end{table}

\begin{table}[H]
\centering
\begin{tabular}{|c|c|c|c|}
\hline
Quarter & Forecast & Actual & Impact \\
\hline
1 & 20.53535 & 20.52920 & -0.006153852 \\
2 & 20.54898 & 20.43943 & -0.109547048 \\
3 & 20.56261 & 20.52514 & -0.037466292 \\
4 & 20.57624 & 20.53204 & -0.044198487 \\
5 & 20.58987 & 20.53768 & -0.052188495 \\
6 & 20.60350 & 20.54956 & -0.053935623 \\
7 & 20.61712 & 20.57584 & -0.041285459 \\
8 & 20.63075 & 20.59461 & -0.036139678 \\
9 & 20.64438 & 20.61742 & -0.026957691 \\
10 & 20.65801 & 20.62636 & -0.031645067 \\
11 & 20.67164 & 20.64051 & -0.031124884 \\
12 & 20.68527 & 20.65443 & -0.030838342 \\
13 & 20.69889 & 20.67194 & -0.026957379 \\
14 & 20.71252 & 20.68736 & -0.025162660 \\
\hline
\end{tabular}
\caption{Impact on the CONS industry}
\label{tab:impact_cons}
\end{table}

\begin{table}[H]
\centering
\begin{tabular}{|c|c|c|c|}
\hline
Quarter & Forecast & Actual & Impact \\
\hline
1 & 20.90446 & 20.89607 & -0.008393570 \\
2 & 20.91308 & 20.84700 & -0.066076739 \\
3 & 20.92170 & 20.88717 & -0.034528975 \\
4 & 20.93032 & 20.91854 & -0.011786348 \\
5 & 20.93894 & 20.91702 & -0.021925376 \\
6 & 20.94756 & 20.93585 & -0.011711461 \\
7 & 20.95619 & 20.93859 & -0.017594648 \\
8 & 20.96481 & 20.96615 & 0.001339304 \\
9 & 20.97343 & 20.98616 & 0.012733796 \\
10 & 20.98205 & 20.99989 & 0.017839575 \\
11 & 20.99067 & 21.03013 & 0.039457200 \\
12 & 20.99929 & 21.03462 & 0.035330132 \\
13 & 21.00791 & 21.04379 & 0.035878698 \\
14 & 21.01653 & 21.05584 & 0.039305078 \\
\hline
\end{tabular}
\caption{Impact on the MAN industry}
\label{tab:impact_man}
\end{table}

\begin{table}[H]
\centering
\begin{tabular}{|c|c|c|c|}
\hline
Quarter & Forecast & Actual & Impact \\
\hline
1 & 20.22889 & 20.25308 & 0.024194725 \\
2 & 20.23765 & 20.20319 & -0.034462225 \\
3 & 20.24642 & 20.24081 & -0.005614076 \\
4 & 20.25519 & 20.25674 & 0.001546734 \\
5 & 20.26396 & 20.25268 & -0.011278539 \\
6 & 20.27272 & 20.27419 & 0.001465980 \\
7 & 20.28149 & 20.30178 & 0.020286354 \\
8 & 20.29026 & 20.33185 & 0.041590435 \\
9 & 20.29903 & 20.35813 & 0.059107222 \\
10 & 20.30779 & 20.35978 & 0.051990800 \\
11 & 20.31656 & 20.40168 & 0.085117917 \\
12 & 20.32533 & 20.40783 & 0.082499261 \\
13 & 20.33409 & 20.43131 & 0.097218425 \\
14 & 20.34286 & 20.44608 & 0.103220092 \\
\hline
\end{tabular}
\caption{Impact on the WHO industry}
\label{tab:impact_who}
\end{table}

\begin{table}[H]
\centering
\begin{tabular}{|c|c|c|c|}
\hline
Quarter & Forecast & Actual & Impact \\
\hline
1 & 20.42273 & 20.43195 & 0.00922234 \\
2 & 20.43100 & 20.37091 & -0.06009800 \\
3 & 20.43928 & 20.47131 & 0.03203222 \\
4 & 20.44755 & 20.49459 & 0.04704262 \\
5 & 20.45583 & 20.50350 & 0.04767609 \\
6 & 20.46410 & 20.54173 & 0.07762984 \\
7 & 20.47237 & 20.54582 & 0.07344234 \\
8 & 20.48065 & 20.58516 & 0.10451716 \\
9 & 20.48892 & 20.56034 & 0.07141526 \\
10 & 20.49719 & 20.56111 & 0.06391646 \\
11 & 20.50547 & 20.58772 & 0.08225573 \\
12 & 20.51374 & 20.58813 & 0.07438660 \\
13 & 20.52201 & 20.61603 & 0.09401104 \\
14 & 20.53029 & 20.61720 & 0.08691329 \\
\hline
\end{tabular}
\caption{Impact on the RET industry}
\label{tab:impact_ret}
\end{table}

\begin{table}[H]
\centering
\begin{tabular}{|c|c|c|c|}
\hline
Quarter & Forecast & Actual & Impact \\
\hline
1 & 20.22889 & 20.25308 & 0.024194725 \\
2 & 20.23765 & 20.20319 & -0.034462225 \\
3 & 20.24642 & 20.24081 & -0.005614076 \\
4 & 20.25519 & 20.25674 & 0.001546734 \\
5 & 20.26396 & 20.25268 & -0.011278539 \\
6 & 20.27272 & 20.27419 & 0.001465980 \\
7 & 20.28149 & 20.30178 & 0.020286354 \\
8 & 20.29026 & 20.33185 & 0.041590435 \\
9 & 20.29903 & 20.35813 & 0.059107222 \\
10 & 20.30779 & 20.35978 & 0.051990800 \\
11 & 20.31656 & 20.40168 & 0.085117917 \\
12 & 20.32533 & 20.40783 & 0.082499261 \\
13 & 20.33409 & 20.43131 & 0.097218425 \\
14 & 20.34286 & 20.44608 & 0.103220092 \\
\hline
\end{tabular}
\caption{Impact on the WHO industry}
\label{tab:impact_who}
\end{table}

\begin{table}[H]
\centering
\begin{tabular}{|c|c|c|c|}
\hline
Quarter & Forecast & Actual & Impact \\
\hline
1 & 20.42273 & 20.43195 & 0.00922234 \\
2 & 20.43100 & 20.37091 & -0.06009800 \\
3 & 20.43928 & 20.47131 & 0.03203222 \\
4 & 20.44755 & 20.49459 & 0.04704262 \\
5 & 20.45583 & 20.50350 & 0.04767609 \\
6 & 20.46410 & 20.54173 & 0.07762984 \\
7 & 20.47237 & 20.54582 & 0.07344234 \\
8 & 20.48065 & 20.58516 & 0.10451716 \\
9 & 20.48892 & 20.56034 & 0.07141526 \\
10 & 20.49719 & 20.56111 & 0.06391646 \\
11 & 20.50547 & 20.58772 & 0.08225573 \\
12 & 20.51374 & 20.58813 & 0.07438660 \\
13 & 20.52201 & 20.61603 & 0.09401104 \\
14 & 20.53029 & 20.61720 & 0.08691329 \\
\hline
\end{tabular}
\caption{Impact on the RET industry}
\label{tab:impact_ret}
\end{table}

\begin{table}[H]
\centering
\begin{tabular}{|c|c|c|c|}
\hline
Quarter & Forecast & Actual & Impact \\
\hline
1 & 20.08419 & 20.07482 & -0.009370581 \\
2 & 20.09860 & 20.01463 & -0.083970333 \\
3 & 20.11301 & 20.07318 & -0.039839512 \\
4 & 20.12743 & 20.10872 & -0.018708576 \\
5 & 20.14184 & 20.11306 & -0.028782966 \\
6 & 20.15626 & 20.16457 & 0.008307266 \\
7 & 20.17067 & 20.20405 & 0.033382259 \\
8 & 20.18509 & 20.23436 & 0.049276182 \\
9 & 20.19950 & 20.25187 & 0.052367600 \\
10 & 20.21392 & 20.25840 & 0.044485946 \\
11 & 20.22833 & 20.29297 & 0.064639808 \\
12 & 20.24274 & 20.29200 & 0.049253502 \\
13 & 20.25716 & 20.31560 & 0.058444174 \\
14 & 20.27157 & 20.32877 & 0.057196374 \\
\hline
\end{tabular}
\caption{Impact on the TRANS industry}
\label{tab:impact_trans}
\end{table}

\begin{table}[H]
\centering
\begin{tabular}{|c|c|c|c|}
\hline
Quarter & Forecast & Actual & Impact \\
\hline
1 & 20.62457 & 20.67137 & 0.04679806 \\
2 & 20.64209 & 20.70152 & 0.05943327 \\
3 & 20.65167 & 20.70140 & 0.04973617 \\
4 & 20.65096 & 20.75058 & 0.09962014 \\
5 & 20.66294 & 20.76529 & 0.10234782 \\
6 & 20.67015 & 20.77214 & 0.10199138 \\
7 & 20.67915 & 20.75090 & 0.07175580 \\
8 & 20.68748 & 20.78957 & 0.10209810 \\
9 & 20.69606 & 20.80174 & 0.10568484 \\
10 & 20.70454 & 20.78463 & 0.08008795 \\
11 & 20.71306 & 20.80946 & 0.09639636 \\
12 & 20.72157 & 20.80067 & 0.07910170 \\
13 & 20.73009 & 20.81664 & 0.08654997 \\
14 & 20.73860 & 20.83563 & 0.09703279 \\
\hline
\end{tabular}
\caption{Impact on the FIN industry}
\label{tab:impact_fin}
\end{table}

\begin{table}[H]
\centering
\begin{tabular}{|c|c|c|c|}
\hline
Quarter & Forecast & Actual & Impact \\
\hline
1 & 19.62940 & 19.63129 & 0.001886092 \\
2 & 19.63935 & 19.52807 & -0.111286751 \\
3 & 19.64931 & 19.63254 & -0.016764850 \\
4 & 19.65926 & 19.67062 & 0.011363035 \\
5 & 19.66921 & 19.72948 & 0.060268903 \\
6 & 19.67917 & 19.74379 & 0.064622668 \\
7 & 19.68912 & 19.77718 & 0.088061295 \\
8 & 19.69907 & 19.80214 & 0.103070442 \\
9 & 19.70903 & 19.81231 & 0.103282509 \\
10 & 19.71898 & 19.78160 & 0.062624764 \\
11 & 19.72893 & 19.76918 & 0.040245071 \\
12 & 19.73889 & 19.75614 & 0.017253893 \\
13 & 19.74884 & 19.76288 & 0.014041773 \\
14 & 19.75879 & 19.76844 & 0.009649179 \\
\hline
\end{tabular}
\caption{Impact on the RE industry}
\label{tab:impact_re}
\end{table}

\begin{table}[H]
\centering
\begin{tabular}{|c|c|c|c|}
\hline
Quarter & Forecast & Actual & Impact \\
\hline
1 & 19.26395 & 19.25966 & -4.288729e-03 \\
2 & 19.27208 & 19.22275 & -4.933148e-02 \\
3 & 19.28197 & 19.23210 & -4.986863e-02 \\
4 & 19.28835 & 19.26050 & -2.784316e-02 \\
5 & 19.29865 & 19.28347 & -1.518380e-02 \\
6 & 19.30895 & 19.30877 & -1.842671e-04 \\
7 & 19.31926 & 19.31912 & -1.400356e-04 \\
8 & 19.32956 & 19.32965 &  8.569209e-05 \\
9 & 19.33987 & 19.32884 & -1.102637e-02 \\
10 & 19.35017 & 19.34350 & -6.666293e-03 \\
11 & 19.36047 & 19.36972 &  9.241764e-03 \\
12 & 19.37078 & 19.38049 &  9.715201e-03 \\
13 & 19.38108 & 19.40759 &  2.650522e-02 \\
14 & 19.39139 & 19.40968 &  1.829096e-02 \\
\hline
\end{tabular}
\caption{Impact on the EDU industry}
\label{tab:impact_edu}
\end{table}

\begin{table}[H]
\centering
\begin{tabular}{|c|c|c|c|}
\hline
Quarter & Forecast & Actual & Impact \\
\hline
1 & 21.11722 & 21.12801 &  0.010791924 \\
2 & 21.12719 & 21.06098 & -0.066206104 \\
3 & 21.13715 & 21.15254 &  0.015388224 \\
4 & 21.14712 & 21.16283 &  0.015711709 \\
5 & 21.15709 & 21.15364 & -0.003451637 \\
6 & 21.16706 & 21.16252 & -0.004535641 \\
7 & 21.17702 & 21.18911 &  0.012087130 \\
8 & 21.18699 & 21.20503 &  0.018036631 \\
9 & 21.19696 & 21.21553 &  0.018577366 \\
10 & 21.20692 & 21.23279 &  0.025864056 \\
11 & 21.21689 & 21.25946 &  0.042565516 \\
12 & 21.22686 & 21.26923 &  0.042375542 \\
13 & 21.23682 & 21.29676 &  0.059940730 \\
14 & 21.24679 & 21.31633 &  0.069542914 \\
\hline
\end{tabular}
\caption{Impact on the HEA industry}
\label{tab:impact_hea}
\end{table}
\begin{table}[H]
\centering
\begin{tabular}{|c|c|c|c|}
\hline
Quarter & Forecast & Actual & Impact \\
\hline
1 & 19.97820 & 19.94579 & -0.0324045231 \\
2 & 19.99339 & 19.37329 & -0.6200994973 \\
3 & 20.00857 & 19.72411 & -0.2844666523 \\
4 & 20.02376 & 19.78110 & -0.2426571383 \\
5 & 20.03895 & 19.81858 & -0.2203678834 \\
6 & 20.05413 & 20.02000 & -0.0341382719 \\
7 & 20.06932 & 20.12043 &  0.0511123425 \\
8 & 20.08451 & 20.11314 &  0.0286351254 \\
9 & 20.09970 & 20.05289 & -0.0468089027 \\
10 & 20.11488 & 20.08619 & -0.0286951382 \\
11 & 20.13007 & 20.12922 & -0.0008478663 \\
12 & 20.14526 & 20.12996 & -0.0152985268 \\
13 & 20.16044 & 20.16836 &  0.0079128986 \\
14 & 20.17563 & 20.18743 &  0.0117986740 \\
\hline
\end{tabular}
\caption{Impact on the ACCF industry}
\label{tab:impact_accf}
\end{table}

\begin{table}[H]
\centering
\begin{tabular}{|c|c|c|c|}
\hline
Quarter & Forecast & Actual & Impact \\
\hline
1 & 21.46856 & 21.47727 &  0.008710057 \\
2 & 21.47835 & 21.45001 & -0.028344818 \\
3 & 21.48766 & 21.47337 & -0.014292772 \\
4 & 21.49680 & 21.47393 & -0.022866274 \\
5 & 21.50587 & 21.48354 & -0.022334524 \\
6 & 21.51491 & 21.49191 & -0.023006661 \\
7 & 21.52395 & 21.50901 & -0.014939644 \\
8 & 21.53298 & 21.50878 & -0.024204846 \\
9 & 21.54201 & 21.51425 & -0.027765223 \\
10 & 21.55104 & 21.51842 & -0.032618243 \\
11 & 21.56007 & 21.52843 & -0.031643781 \\
12 & 21.56910 & 21.54292 & -0.026176581 \\
13 & 21.57813 & 21.55948 & -0.018651811 \\
14 & 21.58716 & 21.57223 & -0.014925834 \\
\hline
\end{tabular}
\caption{Impact on the GOV industry}
\label{tab:impact_gov}
\end{table}

\end{document}